# Genetic Optimization of Keywords Subset in the Classification Analysis of Texts Authorship

## Bohdan Pavlyshenko

*Ivan Franko Lviv National University, Ukraine, pavlsh@yahoo.com*

**Abstract**

The genetic selection of keywords set, the text frequencies of which are considered as attributes in text classification analysis, has been analyzed. The genetic optimization was performed on a set of words, which is the fraction of the frequency dictionary with given frequency limits. The frequency dictionary was formed on the basis of analyzed text array of texts of English fiction. As the fitness function which is minimized by the genetic algorithm, the error of nearest k neighbors classifier was used. The obtained results show high precision and recall of texts classification by authorship categories on the basis of attributes of keywords set which were selected by the genetic algorithm from the frequency dictionary.

**Introduction**

The analysis of texts authorship takes an important place in data mining. In classification and cluster analysis of textual data the vector model of text documents is widely used, where those documents are represented as vectors in some phase space [Pantel and Turney, 2010]. Text frequencies of words form the basis of this space. One of important tasks is to find the optimal vector subspaces of documents for classification and cluster analysis of text documents [Sebastiani, 2002; Forman, 2003]. In particular the problem lies in the selection of keywords, the text frequencies of which can be used as input parameters for text classifiers with satisfactory precision. The solution of this problem optimizes the required number of calculations and the precision of a classifier in data mining. Different parameters are used in the classification analysis for determining the classification potential of each word. However, a set of words may have general synergistic potential which is not visible in every single word.

Genetic algorithms are widely used in the artificial intelligence [Booker et al., 1989; Whitley, 1994]. Genetic optimization can be used for attribute selection in the classification analysis [Vafaie and De Jong, 1992; Raymer et al., 2000; Tan et al., 2008]. Genetic algorithms are also used for text mining [Atkinson-Abutridy, 2004]. In this paper we use the genetic algorithm to select a subset of keywords from chosen range in the frequency dictionary. This keywords subset forms bases of documents vector space in the authorship classification analysis. For our analysis we have chosen texts of English fiction categorized by authors. As the fitness function for genetic optimization we chose the precision of a classifier by *k* nearest neighbors.

**Genetic Optimization of the Basis of Documents Vector Space.**

Genetic algorithms are used in a wide class of optimization problems which consist in finding a set of input parameters that minimize some fitness function. As the fitness function we can consider the classifier's error or some quantitative characteristics of cluster structure that text documents possess. As input parameters of optimization problem we consider a set of keywords that form the basis of text documents vector space. The concept of genetic algorithms consists in using the main principles of Darwin's evolutionary theory, in particular the law of natural selection and genetic variability in optimization problems solving. Let us consider the basic theses of genetic algorithms in the context of the problem of finding an optimal basis for data mining of text documents, in particular on the basis of classification algorithms. In considering the genetic algorithm for finding the optimal keywords basis we use the classical scheme [Booker et al., 1989; Whitley, 1994]. A set of input parameters is called a chromosome or individual. In simple case, an individual is formed on the basis of a single chromosome. A set of chromosomes forms a population. A set of vector basis keywords in the context of genetic algorithms we called a keywords chromosome. A classical genetic algorithm includes the following steps:
1. The initial population of *n* chromosomes is formed.
2. For each chromosome the fitness function is defined.
3. Based on the specified selection rules, two parent chromosomes are selected on the basis of which a new child chromosome for the next population will be formed.
4. For selected parent pairs a crossover operator is applied, by means of which a new child chromosome is formed.
5. The mutation of chromosomes with the given probability is effected.
6. The steps 3-5 are being repeated until a new population of *n* chromosomes is generated.
7. The steps 2-6 are being repeated until they meet conditions of algorithm stop. Such a condition can be, for example, an assigned set of the fitness function value, or the maximum number of iterations.

In discrete optimization with the use of genetic algorithms, the number of steps required to find the optimal set of input parameters is polynomially less as compared to the enumeration of possibilities. This is due to the presence of some sections of chromosomes which are somewhat similiar to genes by their behavior and which collectively make optimization contribution to the fitness function. That means that input parameters are considered as some groups (genes) that chromosomes are being exchanged by, using a crossover operator, which reduces significantly the number of parameters combinations in the optimization analysis.

Let us consider a set-theoretical model of the genetic algorithm of optimization selection semantic fields for forming the semantic space of text documents. We consider the evolution of genetic optimization as an ordered set of populations

$$Ev = \{Pop_k \mid k = 1,2,...\mid Ev\mid\}. \quad (1)$$

We assume that one generation of chromosomes is formed by one population. The population consists of a set of chromosomes

$$Pop_k = \{\chi_{jk}^p \mid j = 1,2,...\mid Pop_k\mid; k = 1,2,...\mid Ev\mid\}. \quad (2)$$

Generally different populations may contain a different number of chromosomes. In simplified case, we suppose that the number of chromosomes is the same in all populations, i. e.

$$\mid Pop_k \mid = \mid Pop \mid = N_{pop}^{\chi}. \quad (3)$$

We consider each chromosome as a set of keywords

$$\chi_{jk} = \{w_{ijk}^{xp} \mid i = 1,2...\mid \chi\mid; j = 1,2,...\mid Pop\mid; k = 1,2,...\mid Ev\mid\}, \quad (4)$$

Where $i$ is the index of the keyword position in the chromosome $\chi_{jk}$ of the population $Pop_k$. Text documents are represented as vectors of keywords text frequencies $p_{kj}^{wd}$ that mean the frequency of the keyword $w_j$ in the text document $d_j$. The set of values $p_{kj}^{wd}$ form the feature-document matrix where the features are the keywords frequencies in the documents:

$$M_{wd} = \left(p_{kj}^{wd}\right)_{k=1, j=1}^{N_s, N_d}. \quad (5)$$

The vector

$$V_j^w = \left(p_{1j}^{wd}, p_{2j}^{wd},..., p_{N_w j}^{wd}\right)^T \quad (6)$$

displays the document $d_j$ in $N_w$–dimensional space of text documents with the basis formed by keywords. Now we consider the use of the genetic algorithm for the optimization of keywords set in the task of text documents classification. The words with the largest text frequencies carry the minimal semantic information, so it is important to choose such set of words for genetic selection of keywords, which will consist of the words that carry the semantics of a text. Such words in the structure of the frequency dictionary are those of medium and minimum frequencies. As an initial set of attributes we consider some fraction of a frequency dictionary with given frequencies limits

$$W_g = \{w_i \mid w_i \in W_f, p_{min}^f \leq p(w_i) < p_{max}^f \}, \qquad (7)$$

where $W_f$ is a set of words of the frequency dictionary; $p_{min}^f$, $p_{max}^f$ are the minimum and maximum limits of the frequency dictionary fraction. As a fitness function for evolutionary optimization of the keywords set of the vector space basis we examine the precision of the classifier. Suppose there are some categories of text documents. These categories may have different nature, for example, they can identify author's idiolect, discourse, or characterize different objects, phenomena, events, etc. We denote the set of these categories as

$$Categories = \{Ctg_m \mid m = 1,2,...,N_{ctg} \}, \qquad (8)$$

where $N_{ctg} = |Categories|$ defines the size of categories set. According to given categories the text documents of the document set $D$ are distributed. The task is to find the fitness function that is described by the mapping

$$F_{d \to ctg} : Categories \times D \to \{0,1\}. \qquad (9)$$

The precision characteristic is widely used to characterize classifiers. The precision of the classifier for the category $Ctg_j$ is defined as the ratio of the number of items, correctly classified as belonging to the category $Ctg_j$, to the total number of items, classified as belonging to the category $Ctg_j$

$$\Pr_j = \frac{|\{d_i \mid Class(d_i) = Ctg_j \wedge d_i \in Ctg_j\}|}{|\{d_i \mid Class(d_i) = Ctg_j\}|}, \qquad (10)$$

where $Class(d_i)$ is the category defined by the classifier. Let us define the fitness function of genetic optimization as follows:

$$F_s^{ga} = 1 - \Pr_{avg}, \qquad (11)$$

where $\Pr_{avg}$ is the classifier's precision averaged by all categories. The target of genetic optimization is to minimize the fitness function $F_s^{ga}$.

As the classification method in the study of genetic optimization we consider the classification by the nearest $k$ neighbors that is called the kNN classification [Sebastiani, 2002; Manning et al., 2008]. This method is referred to as vector classifiers. The basis of vector classification methods is the hypothesis of compactness. According to this hypothesis, the documents belonging to one and the same class form a compact domain, and the domains that belong to different classes do not intersect. As a similarity measure between the documents we chose Euclidean distance. In kNN classification the boundaries of categories are defined locally. Some documents are referred to a category which is dominant for its $k$ neighbors. In the case $k = 1$ the document obtains the category of its nearest neighbor. Due to the compactness hypothesis a test document $d$ has the same category as most of the documents in the training sample in some local spatial neighborhood of a document *d*. In the genetic selection of semantic fields we use the indices of keywords as input parameters of the optimization problem. The result of genetic optimization will be a set of indices which determines the optimal set of keywords.

**Experimental part**

The experimental array of text documents consisted of 503 English fiction texts that were classified by the categories of 17 authors. The study sample consisted of 300 randomly selected documents, and the test sample consisted of 153 documents. The set of keywords for genetic optimization is formed by the first 1000 words of the frequency dictionary which have the text frequency less than 0.001. These words form a frequency interval $[7.70 \cdot 10^{-5}, 10^{-3}]$. The populations of 50 chromosomes size were under analysis. The operator of scattered crossover with the size fraction 0.8 was applied. In each population five elite chromosomes were selected. The chromosomes with the size of 30 and 10 keywords were under analysis. Fig. 1 shows the dynamics of minimum fitness function value and the one averaged over the populations at the chromosome size of 30. The classifier by the *k* nearest neighbors was chosen for calculating the fitness function which is based on the classification precision. The resulting minimum fitness function value is 0.0858. This fitness function value corresponds to the following set of keywords:

*{name, seven, together, mind, meeting, north, threw, laid, fifty, rate, cast, move, blow, took, showed, opinion, make, shook, leave, feel, times, address, around, chief, next, hall, half, tea, worth, started}*

Obtained keywords set was used for the classification texts by authorship categories. As an additional classifier characteristics the recall $Rc_j$ is used, which is defined as the ratio of the number of items, correctly classified as belonging to the category $Ctg_j$, to the total number of items, which belong to the category $Ctg_j$.

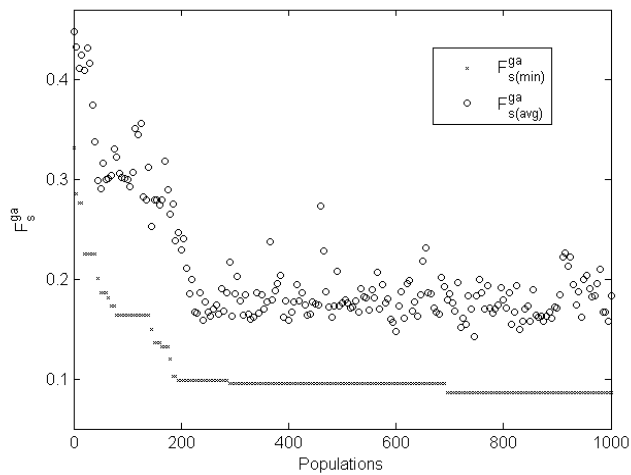

*Fig.1 Dynamics of minimum $F_{s(\min)}^{ga}$ and average $F_{s(avg)}^{ga}$ values for the fitness function at the chromosome size of 30.*

Fig. 2 shows the diagram of classifier's precision and recall by the nearest neighbors for the obtained set of 30 words which corresponds to the minimum value of fitness function. Obtained precision and recall, averaged over authorship categories, are the following: $Pr = 0.8417$, $Rc = 0.8220$. In case of random formation of training and test samples, the distribution precision and recall by authors' categories will vary with each application of the classifier to the test sample. However, the main parameters such as the average values of precision and recall will be similar.

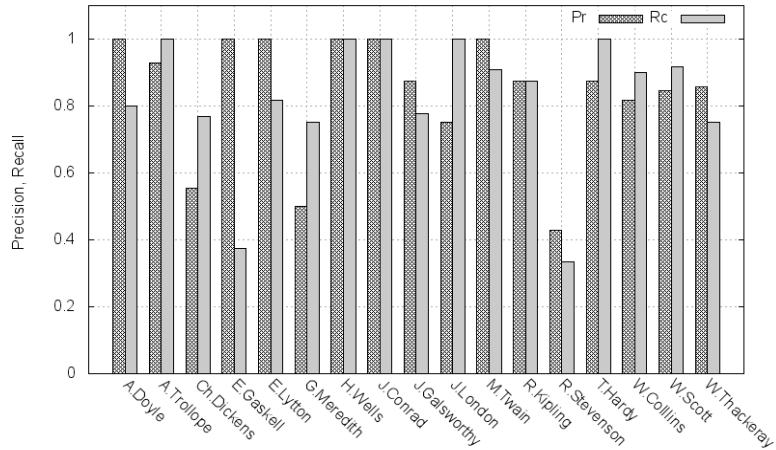

*Fig.2 Classifier's precision Pr and recall Rc for the authorship categories at optimized set of 30 words.*

We also used the genetic optimization of keywords set, when the size of the chromosome was 10 keywords. In this case the dimension of documents vector space was 10. Fig. 3 shows the dynamics of minimum and averaged over the population fitness function value at the chromosome size of 10. The resulting minimum fitness function value is equal to 0.1923.

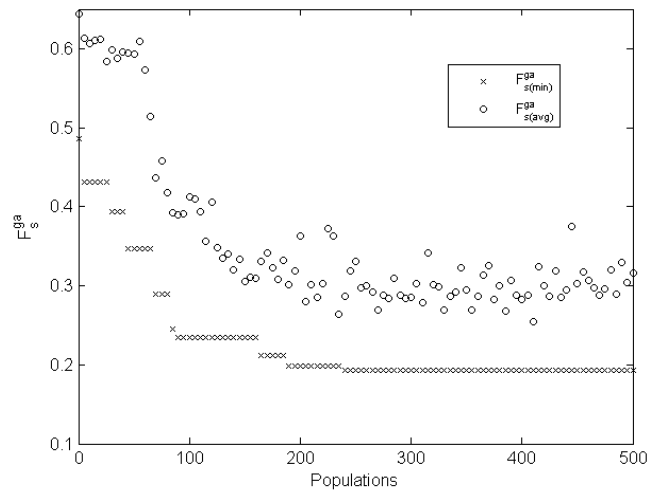

*Fig.3. Dynamics of minimum $F_{s(\min)}^{ga}$ and average $F_{s(avg)}^{ga}$ values for the fitness function at the chromosome size of 10.*

Fig. 4 shows a diagram of the classifier's precision and recall by the nearest neighbors for the obtained set of 10 words. Obtained precision and recall, averaged over authorship categories, are the following: $Pr = 0.7566$, $Rc = 0.7652$. Comparing the data obtained for the chromosomes with the size of 30 and 10 keywords, one can see that taking a third of the space dimension, the precision and recall decreased by only 10%.

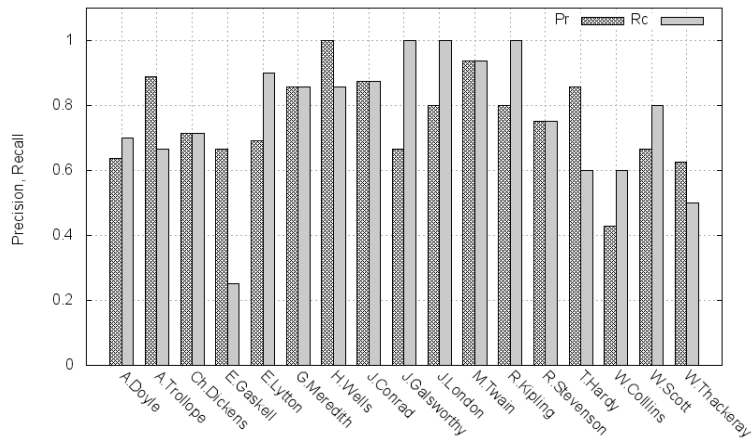

*Fig.4. Classifier's precision Pr and recall Rc for the authorship categories at optimized set of 10 words.*

**Summary and Conclusions**

The paper described the genetic optimization of keywords, the frequencies of which are the components of documents vectors and they act as attributes in text classification analysis. The genetic optimization was performed on the set of words, which is the fraction of the frequency dictionary with given frequency limits. The frequency dictionary was formed on the basis of analyzed text array of texts of English fiction. As the fitness function which is minimized by the genetic algorithm, the error of nearest $k$ neighbors classifier was used. The obtained results show high precision and recall of texts classification by authorship categories on the basis of multiple attributes of keywords that were selected by the genetic algorithm from the frequency dictionary.